# Competing effects at Pt/YIG interfaces: spin Hall magnetoresistance, magnon excitations and magnetic frustration

Saül Vélez[1,*], Amilcar Bedoya-Pinto[1,‡], Wenjing Yan[1], Luis E. Hueso[1,2], and Fèlix Casanova[1,2,†]

[1]CIC nanoGUNE, 20018 Donostia-San Sebastian, Basque Country, Spain
[2]IKERBASQUE, Basque Foundation for Science, 48011 Bilbao, Basque Country, Spain

[‡] Present address: Max Planck Institute of Microstructure Physics, D-06120 Halle, Germany
[*] s.velez@nanogune.eu
[†] f.casanova@nanogune.eu

We study the spin Hall magnetoresistance (SMR) and the magnon spin transport (MST) in Pt/$Y_3Fe_5O_{12}$(YIG)-based devices with intentionally modified interfaces. Our measurements show that the surface treatment of the YIG film results in a slight enhancement of the spin-mixing conductance and an extraordinary increase in the efficiency of the spin-to-magnon excitations at room temperature. The surface of the YIG film develops a surface magnetic frustration at low temperatures, causing a sign change of the SMR and a dramatic suppression of the MST. Our results evidence that SMR and MST could be used to explore magnetic properties of surfaces, including those with complex magnetic textures, and stress the critical importance of the non-magnetic/ferromagnetic interface properties in the performance of the resulting spintronic devices.

**I. Introduction**

*Insulating spintronics* [1] has emerged as a promising, novel technological platform based on the integration of ferromagnetic insulators (FMIs) in devices as a media to generate, process and transport spin information over long distances [1–30]. The advantage of using FMIs against metallic ones is that the flow of charge currents is avoided, thus preventing ohmic losses or the emergence of undesired spurious effects. Some phenomena explored in insulating spintronics include the spin pumping [2–5], the spin Hall magnetoresistance (SMR) [5–15], the spin Seebeck effect [5,16–18], the spin Peltier effect [19], the magnetic gating of pure spin currents [20,21] or the magnon spin transport (MST) [2,22–30].

The fundamental building block structure employed to explore these phenomena is formed by a FMI layer –typically $Y_3Fe_5O_{12}$ (YIG) due to its small damping, soft ferrimagnetism and negligible magnetic anisotropy– and a non-magnetic (NM) metal with strong spin-orbit coupling (SOC) such as Pt or Ta placed next to it, which is essentially used to either generate or detect spin currents *via* the spin Hall effect (SHE) or its inverse [31–35]. Since these spintronic phenomena are based on the transfer of spin currents across the NM/FMI interface, it plays a key role in the properties and the performance of the resulting devices.

It is well established that the most relevant parameter that determines the spin-current transport across the interface is the spin-mixing conductance $G_{\uparrow\downarrow} = (G_r + iG_i)$ [5,36,37]. However, it is still under debate whether other interface effects could also be relevant in these hybrid systems. Some examples are the magnetic proximity effect (MPE) [38–43], the Rashba-Edelstein effect [44–47], the anomalous Nernst effect [38,48,49] or the spin-dependent interfacial scattering [50]. Therefore, understanding the role of the NM/FMI interface and the impact of its properties on the resulting spintronic phenomena is of outmost importance.

In this work, we show that different spin-dependent phenomena in Pt/YIG-based devices (SMR and MST) are dramatically altered when the YIG surface is treated with a soft Ar$^+$-ion milling. At room temperature, while the SMR effect in the treated samples is slightly larger than in the non-treated ones, the MST signal is fourfold increased. This extraordinary increase in the MST amplitude indicates that the spin-to-magnon conversion in Pt/YIG interfaces is strongly dependent on the magnetic details of the atomic layer of the YIG beyond the change in $G_{\uparrow\downarrow}$. In addition, at low temperature, we observe a sign change of the SMR and a strong suppression of the MST signal in the treated samples, indicating the emergence of a surface magnetic frustration of the treated YIG at low temperature. Our experimental results point out SMR and MST to be powerful tools to explore magnetic properties of surfaces and show that care should be taken when treating the surface of YIG, especially when used for studying spin-dependent phenomena originating at interfaces.

**II. Experimental details**

Two different types of device structures were studied. In the first design, Pt/YIG samples were prepared by patterning a Pt Hall bar (width $W$=100 µm, length $L$=800 µm and thickness $d_N$=7 nm) on top of a 3.5-µm-thick YIG film [51] *via* e-beam lithography, sputtering deposition of Pt and lift-off, as fabricated in Ref. 52. In some samples, the YIG top surface was treated with a gentle Ar$^+$-ion milling [53] prior the Pt deposition (Pt/YIG$^+$ samples). In the second design, non-local NL-Pt/YIG and NL-Pt/YIG$^+$ lateral nanostructures were prepared on top of a 2.2-µm-thick YIG film [51] by patterning two long Pt strip lines ($W$=300 nm, $L_1$=15.0 µm, $L_2$=12.0 µm and $d_N$=5 nm) separated by a gap of ~500 nm –similar to the device structure used in Refs. 25 and 29–, following the same fabrication procedure used for the Hall bar. For each device structure, the Pt for both treated and non-treated samples was deposited in the same run. Here, for the sake of clarity, we present data taken for one sample of each type (Pt/YIG, Pt/YIG$^+$, NL-Pt/YIG and NL-Pt/YIG$^+$), although more samples were fabricated and measured, all showing reproducible results.

Magnetotransport measurements were performed using a Keithley 6221 sourcemeter and a Keithley 2182A nanovoltmeter operating in the dc-reversal method [54–56]. These measurements were performed at different temperatures between 10 and 300 K in a liquid-He cryostat that allows applying magnetic fields $H$ of up to 9 T and to rotate the sample by 360º degrees. No difference in the magnetic properties between YIG and YIG$^+$ substrates were observed *via* VSM magnetometry measurements (not shown).

**III. Results and Discussion**

**IIIa. Spin Hall magnetoresistance**

First, we explore the angular-dependent magnetoresistance (ADMR) in Pt/YIG and Pt/YIG$^+$ at room temperature. Figures 1(a)-1(c) show the longitudinal ($R_L$) ADMR curves obtained for both samples in the three relevant **H**-rotation planes. The transverse ($R_T$) ADMR curves taken in the $\alpha$ plane are plotted in Fig. 1(d). The measurement configuration, the definition of the axes, and the rotation angles ($\alpha, \beta, \gamma$) are defined in the sketches next to each panel. Note that for the magnetic fields applied, the magnetization of the YIG film is saturated [see Ref. 52 for the characterization of the YIG films]. The angular dependences are the same in both milled and non-milled samples and show the expected behaviour for the SMR effect, in agreement with measurements reported earlier in Pt/YIG bilayers [5–7,11,52].

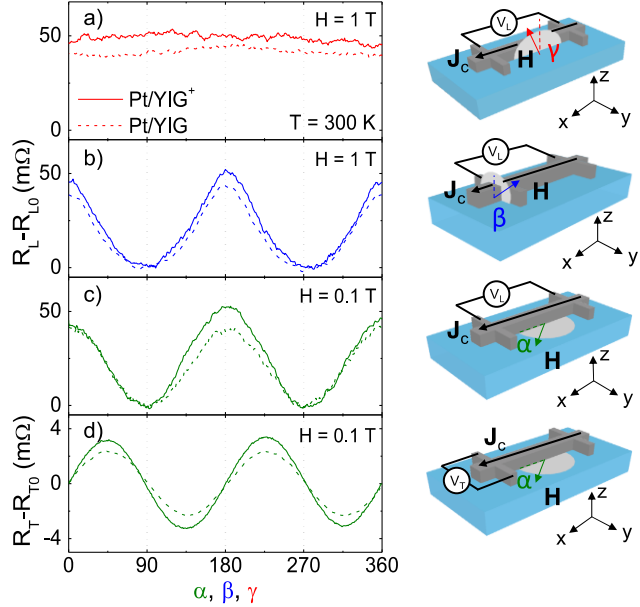

FIG. 1 (color online). (a)-(c) Longitudinal ADMR measurements performed in Pt/YIG (dashed lines) and Pt/YIG[+] (solid lines) samples at 300 K in the three relevant **H**-rotation planes ($\alpha, \beta, \gamma$). (d) Transverse ADMR measurements taken in the same samples and temperature in the $\alpha$ plane. Sketches on the right side indicate the definition of the angles, the axes, and the measurement configuration. The applied magnetic field is denoted in each panel. $R_{L0}$ and $R_{T0}$ are the subtracted base resistances.

The SMR arises from the interaction of the spin currents generated in the NM layer due to the SHE with the magnetic moments of the FMI. According to the SMR theory [8,52], the longitudinal and transverse resistivities of the Pt layer are given by

$$\rho_L = \rho_0 + \Delta\rho_0 + \Delta\rho_1 (1 - m_y^2),$$
$$\rho_T = \Delta\rho_1 m_x m_y + \Delta\rho_2 m_z, \qquad (1)$$

where $\mathbf{m}(m_x, m_y, m_z) = \mathbf{M}/M_s$ are the normalized projections of the magnetization of the YIG film to the three main axes, $M_s$ is the saturated magnetization of the YIG and $\rho_0$ is the Drude resistivity. $\Delta\rho_0$ accounts for a number of corrections due to the SHE [52,57,58], $\Delta\rho_1$ is the main SMR term, and $\Delta\rho_2$ accounts for an anomalous Hall-like contribution. Considering that these magnetoresistance (MR) corrections are very small, we identify the base resistivity of our longitudinal ADMR measurements as $\rho_{L0}(m_y = 1) = \rho_0 + \Delta\rho_0 \simeq \rho_0$. Since **H** is rotated in the plane of the film in our transverse measurements, the $\Delta\rho_2$ contribution does not appear. Note that, in ADMR measurements, the amplitude of $\rho_L(\beta)$, $\rho_L(\alpha)$ and $\rho_T(\alpha)$ are equal and given by $\Delta\rho_1$. Therefore, these measurements are equivalent when only the SMR contributes to the MR. The SMR term is quantified by

$$\frac{\Delta\rho_1}{\rho_0} = \theta_{SH}^2 \frac{\lambda}{d_N} \text{Re} \frac{2\lambda G_{\uparrow\downarrow}\rho_0\tanh^2(d_N/2\lambda)}{1+2\lambda G_{\uparrow\downarrow}\rho_0\coth(d_N/\lambda)}, \quad (2)$$

where $\lambda$ is the spin diffusion length and $\theta_{SH}$ the spin Hall angle of the Pt layer.

According to Eqs. (1) and (2), the difference in the SMR amplitude observed between the two samples (see Fig. 1) can be interpreted as an enhanced $G_{\uparrow\downarrow}$ at the Pt/YIG[+] interface –with respect to Pt/YIG– due to the Ar[+]-ion milling process. Note that the spin transport properties for both Pt layers are expected to be the same because the measured resistivity is the same [59–61]. As the spin relaxation is governed by the Elliott-Yafet mechanism in Pt [59–61], we can calculate its spin diffusion length using the relation $\lambda = (0.61\times10^{-15}\Omega m^2)/\rho$ [61]. Following Ref. 61, the spin Hall angle in the moderately dirty regime can be calculated using the intrinsic spin Hall conductivity $\sigma_{SH}^{int}$ ($\theta_{SH} = \sigma_{SH}^{int}\rho$), which for Pt is 1600 $\Omega^{-1}cm^{-1}$ [61,62]. In our films, $\rho_{L0} \sim 63$ μΩcm at 300 K, which thus corresponds to $\lambda\sim1.0$ nm and $\theta_{SH}\sim0.097$. Using these $\lambda$ and $\theta_{SH}$ values, $d_N=7$ nm, $\Delta\rho_L/\rho_0\sim5.3\times10^{-5}$ and $\sim7.06\times10^{-5}$ (for Pt/YIG and Pt/YIG[+], respectively, at 300 K), and that $G_i<<G_r$ [63], Eq. (2) yields $G_r \sim3.3\times10^{13}$ $\Omega^{-1}m^{-2}$ and $\sim4.4\times10^{13}$ $\Omega^{-1}m^{-2}$ for the Pt/YIG and Pt/YIG[+] samples, respectively, which is within the range of values reported using the same bilayer structure [2,5–7,9–11,52,64,65]. This increase in $G_r$ is in agreement with previous studies, where it was shown that an Ar[+]-ion milling process can improve the NM/YIG interface quality by removing residues that might remain over the YIG substrate before the deposition of the NM layer [65,66]. However, it has been observed that an Ar[+]-ion milling process might also affect the YIG structure [49,64]. In the following, we proceed to study the temperature dependence of the SMR effect in these samples.

Figures 2(a) and 2(b) show the measured temperature dependence of $R_T(\alpha)$ for Pt/YIG and Pt/YIG[+], respectively, in the angular range 0-180º and for $H$=0.1 T. In both samples, the angular dependence predicted by the SMR effect is preserved when decreasing the temperature, following a $\sin(\alpha)\cos(\alpha)$ dependence [see Eq. (1)]. However, the polarity of the ADMR amplitude reverses the sign for Pt/YIG[+] at low temperatures (crossing zero around $T\sim45$ K), which is a completely unexpected behavior. According to the SMR theory, this amplitude is given by the term $\Delta\rho_1/\rho_0$ in Eq. (2), which is a positive magnitude by definition.

In Fig. 2(c), we plot the temperature dependence of the normalized amplitude of the transverse ADMR $\Delta\rho_T/\rho_0\approx\Delta\rho_T/\rho_{L0}=[[R_T(45º)-R_T(135º)]/R_{L0}]\cdot[L/W]$ for Pt/YIG (black squares) and Pt/YIG[+] (red circles). The weak temperature dependence of the SMR effect observed in our Pt/YIG sample is very similar to the one reported by others using the same bilayer structure and it can be well understood with the temperature evolution of the spin transport properties in Pt [13,14,59,61]. In contrast, the different temperature dependence observed in Pt/YIG[+] [see red dashed line in Fig. 2(c), which shows a scaling of the MR measured in Pt/YIG], having a sharp drop below 140 K and even a sign change at low temperatures, suggests the emergence of an additional interface effect. Systematic ADMR measurements are required to address its origin.

Figure 2(d) shows the temperature dependence of the normalized amplitude of the longitudinal ADMR $\Delta\rho_L/\rho_0\approx\Delta\rho_L/\rho_{L0}=[R_L(0º)-R_L(90º)]/R_{L0}$ measured in Pt/YIG[+] for the three relevant **H**-rotation planes at $H$=1 T. We can see that both $\Delta\rho_L(\alpha)/\rho_0$ and $\Delta\rho_L(\beta)/\rho_0$ follow the same trend and that $\Delta\rho_L(\gamma)/\rho_0$ remains zero, except for T~10 K. At very low temperatures, weak anti-localization effects emerge in Pt thin films [52,67–69], resulting in an extra out-of-plane vs in-plane MR, giving an explanation for the very small signal detected at 10 K. These measurements show that the sudden drop and the change in sign of the MR observed in Pt/YIG[+] when decreasing temperature preserve the symmetry given by the polarization (**s**) of the spin current produced in the Pt layer *via* the SHE, i.e., the measured MR has the symmetry of the SMR effect, which is distinct to the anisotropic MR that would appear if MPE were present. Therefore, this excludes MPE to be at the origin of the sign change of the MR at low temperatures in Pt/YIG[+].

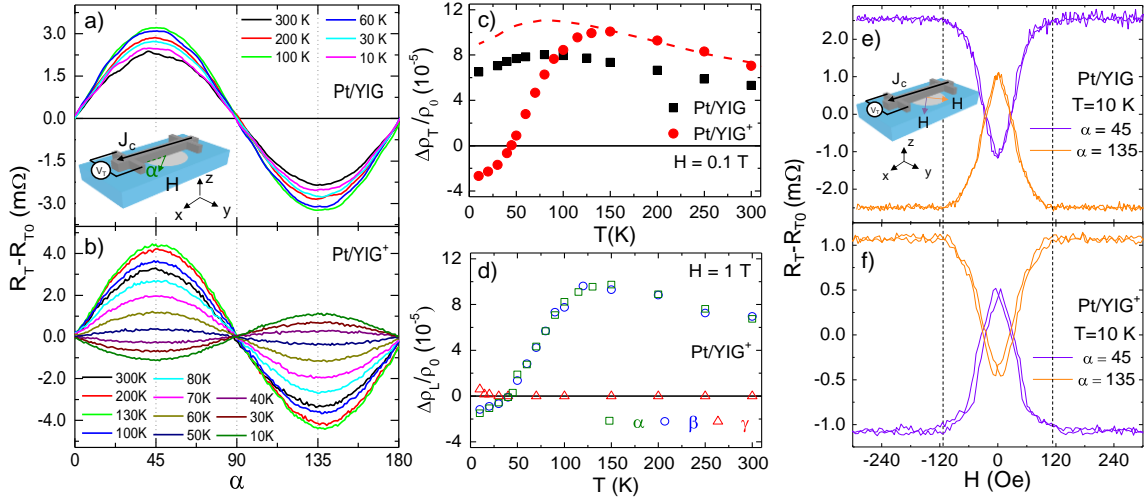

FIG. 2 (color online). (a), (b) Transverse ADMR curves measured in Pt/YIG and Pt/YIG[+], respectively, at different temperatures for $H$=0.1 T in the $\alpha$ plane (see sketch). Data in the 180º-360º range reproduce the same curves. $R_{T0}$ is the subtracted base resistance at the corresponding temperature. (c) Temperature dependence of the normalized amplitude of the transverse ADMR, $\Delta\rho_T/\rho_0$, for the Pt/YIG (black squares) and Pt/YIG[+] (red circles) samples extracted from (a) and (b), respectively. The red dashed line in (c) is a scaling of the temperature dependence of the amplitude measured in Pt/YIG to overlap with the amplitude obtained in Pt/YIG[+] in the high temperature range (from ~150 to 300 K). (d) Temperature dependence of the normalized amplitude of the longitudinal ADMR, $\Delta\rho_L/\rho_0$, obtained in Pt/YIG[+] at $H$=1 T and for the three **H**-rotation planes ($\alpha,\beta,\gamma$). (e), (f) Transverse magnetic-field-dependent MR curves measured in Pt/YIG and Pt/YIG[+], respectively, at 10 K with $H$ in the plane of the film and for $\alpha$=45º and $\alpha$=135º [see sketch in (e) for the color code of the magnetic field direction]. The vertical dashed lines show the saturation field of the YIG film obtained *via* magnetometry measurements.

It is important to point out that, in hybrid systems of this kind, the interaction of **s** with the magnetization **M** of the FMI leads to a resistance modulation not only due to the SMR, but also due to the excitation of magnons [25,29]. While the amplitude of the SMR is maximum when **s** and **M** are perpendicular, the resistance modulation due to magnon excitation is maximized when **s** and **M** are collinear. This implies that the MR modulation obtained in NM/FMI hybrids *via* ADMR measurements must actually be the result of the competition of these two spin-dependent MR effects, having the same angular dependences, but with reversed polarity. However, the MR expected from magnon excitations is much smaller than from the SMR for the range of temperatures explored here. It has been estimated to be ~16 % at room temperature with respect to the SMR [19,21,25], and that it should vanish at zero temperature [29]. Therefore, this rules out the excitation of magnons as responsible for the unexpected MR measured in Pt/YIG$^+$ at low temperatures [see Fig. 2(b) and 2c)]. However, note that the excitation of magnons may lead to a larger correction in the ADMR amplitude at very high temperatures. This could give an alternative explanation to the measured temperature dependence of the MR in Pt/YIG bilayers close to the Curie temperature of the YIG film [15].

Figures 2(e) and 2(f) show the magnetic-field-dependent MR curves measured in Pt/YIG and Pt/YIG$^+$, respectively, at 10 K with the magnetic field applied in the plane of the film and along two representative directions ($\alpha$=45º and $\alpha$=135º). The peaks and dips correspond to the magnetization reversal of the YIG film as reported earlier [6,9–11]. Note that the saturation field of the YIG film obtained via magnetometry measurements (denoted as vertical dashed lines) matches perfectly with the one obtained through MR measurements in both samples. Moreover, the signs of the MR signals (for $\alpha$=45º and $\alpha$=135º) are reversed in Pt/YIG$^+$ with respect to the ones measured in Pt/YIG, which is in agreement with the sign change observed in the ADMR at low temperatures [see Figs. 2(b) and 2(c)].

Because the SMR effect is basically sensitive to the magnetic properties of the first magnetic layer, having an estimated penetration depth of just a few Å [36], all previous measurements indicate that the magnetic moments of the surface of the YIG$^+$ film are perpendicularly coupled to the ones of the bulk at low temperatures. The emergence of this surface magnetic frustration in our treated samples could be caused by a competing ferromagnetic and antiferromagnetic coupling of the modified complex stoichiometry of the YIG film due to the Ar$^+$-ion milling process. In fact, magnetic frustration has already been observed in some ferrimagnets at low temperatures [70–73]. The angle $\phi$ between the magnetic moments of the surface and the bulk magnetization would be maximum (up to 90º) at low temperatures. The fact that the external magnetic field **H** aligns the bulk **M** but the SMR is sensitive to the magnetic moments of the surface yields a negative amplitude of the ADMR. A rise in the temperature would lead to a reduction of the angle $\phi$ due to the increase of the thermal energy in the magnetically coupled system. Considering our measurements, both surface and bulk magnetizations would lie together above ~140 K, recovering the expected positive amplitude of the ADMR.

According to this physical picture, when the magnetic field (with $H>H_S$) rotates in a particular **H**-rotation plane, the magnetic frustration forces the surface magnetization to point to a perpendicular direction. Due to the degeneracy in the orientation where the surface magnetization could point to, the angular dependences of the ADMR signals are preserved. As for the magnetic-field-dependent MR curves, when $H<H_S$, our YIG bulk film breaks in domains [74–76], resulting in the peaks and dips observed [see Fig. 2(e)]. The fact that the estimated $H_S$ of the surface magnetization *via* MR measurements is the same for both samples [see Figs. 2(e) and 2(f)] and correlates with the measured $H_S$ of the film indicates that the magnetic moments of the surface of the YIG$^+$ must be coupled to the bulk. The fact that the peaks and dips in the MR curves are reversed confirms that the angle $\phi$ between the magnetizations of the frustrated surface and the bulk should approach 90º at very low temperatures.

In this scenario, one may think that, by applying a large enough magnetic field, we should be able to exert enough canting to the frustrated surface magnetic moments to shift the ADMR amplitude to positive values (i.e., reduce $\phi$). Positive ADMR values have actually been measured for H>2T at low temperatures. However, the large Hanle magnetoresistance (HMR) effect [52] present in our samples (the measured HMR amplitude at 300 K and 9 T is $\Delta\rho_L/\rho_0 \sim 16\cdot10^{-5}$) dominates the MR at large fields, preventing us from quantifying the canting exerted to the frustrated magnetic moments *via* MR measurements.

An alternative interpretation of the temperature dependence of the SMR, motivated by the results obtained exploring a Pt/NiO/YIG system [77], is that the magnetic moments of the treated YIG$^+$ surface are perpendicularly coupled to the magnetization of the YIG film at any temperature. In this situation, the frustrated magnetization of the surface dominates the SMR at low temperature, which is negative. When increasing the temperature, the frustrated surface becomes more transparent to the spin currents due to the thermal fluctuations and the YIG magnetization progressively dominates the SMR, which becomes positive. In other words, the spin current generated by the Pt reaches the bulk YIG and the usual SMR in Pt/YIG is detected. This competition would lead to a decrease in the SMR amplitude below ~140 K, a compensation at an intermediate temperature (i.e., zero SMR amplitude, which occurs around 45 K in our system), and a negative amplitude at low temperatures, when the frustrated Pt/YIG$^+$ interface dominates.

Our model allows us to qualitatively show that the emergence of a surface magnetic frustration can be well identified *via* SMR measurements. Note that magnetic frustration at the first atomic layer of a film cannot be detected by means of standard surface techniques such as magneto-optical Kerr effect, magnetic force microscopy, or X-ray magnetic circular dichroism because of the relatively long penetration depth. Other surface sensitive techniques such as spin-polarized scanning tunneling microscopy or scanning electron microscopy with polarization analysis cannot be used in magnetic insulators either. Only complex, depth sensitive techniques such as polarized neutron reflectometry might resolve the surface magnetization independently of bulk. In other words, the magnetic properties of the very first layer of an insulating film will generally remain hidden by the large magnetic response of its bulk. Remarkably, unlike other techniques, the SMR can be applied to FMI films, is sensitive to only the first atomic layer [36], and its response is associated to the relative direction of the magnetic moments of the FM with respect to the spins of the NM layer (whether they are parallel or perpendicular), but not to their orientation (up or down). This highlights the potential of the SMR to explore complex surface magnetic properties [78].

**IIIb. Magnon spin transport**

We now move to study the magnon spin transport in the non-local NL-Pt/YIG and NL-Pt/YIG$^+$ samples. Figure 3(a) shows an optical image of one of the devices fabricated. In these samples, the current is injected in the central wire and both the local resistance ($R_L=V_L/I$) and the non-local resistance ($R_{NL}=V_{NL}/I$) are measured as schematically drawn in Fig. 3(a). Note that $R_{NL}$ is measured using the dc-reversal method [54–56], which is equivalent to the first harmonic signal in ac lock-in type measurements [79].

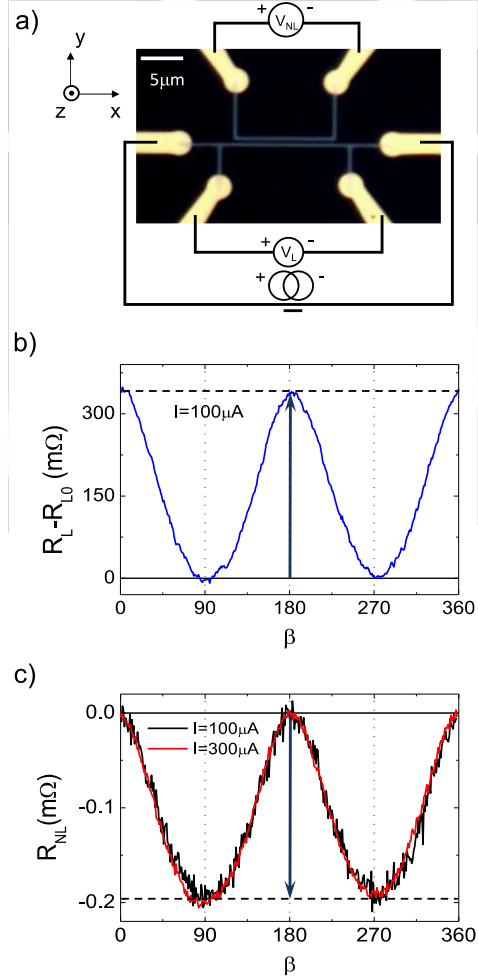

FIG. 3 (color online). (a) Optical image of the NL-Pt/YIG sample. Grey wires are the Pt stripes and the yellow areas correspond to additional Au pads. The black background is the surface of the YIG film. Both the local and non-local measurement configurations are schematically shown. (b) and (c) are the local ($R_L$) and non-local ($R_{NL}$) ADMR signals, respectively, measured in the NL-Pt/YIG sample at 150 K and for $H$=1 T rotating in the $\beta$ plane. Note that, along this rotation angle, **M** changes its relative orientation with **s** (being parallel for $\beta$=90º and 270º and perpendicular for $\beta$=0º and 180º). In (b), the bias current was 100 µA. In (c), non-local ADMR measurements performed at $I$=100 (black line) and 300 µA (red line) are shown. The arrows in (b) and (c) indicate the sign convention used for the amplitude of the local ($\Delta R_L$) and non-local ($\Delta R_{NL}$) resistance plotted in Figs. 4(a) and 4(b), respectively.

Figures 3(b) and 3(c) show an example of the local and non-local ADMR measurements, respectively, performed in our samples. The data correspond to the NL-Pt/YIG sample measured at 150 K with $H$=1 T rotating in the $\beta$ plane [see Fig. 1(b) for the definition]. Similar ADMR curves were obtained in the NL-Pt/YIG$^+$ sample. The local resistance $R_L$ [Fig. 3(b)] shows the expected $\cos^2(\beta)$ dependence for the SMR effect. Taking into account that in these samples $\rho_{L0}$ (300 K)~ 54 $\mu\Omega$ cm –which according to Ref. 61 corresponds to $\lambda$~1.2 nm and $\theta_{SH}$~0.083 for the Pt film–, that the measured SMR amplitudes at the same temperature are $\Delta\rho_L/\rho_0$~6.2×10$^{-5}$ and ~7.6×10$^{-5}$ (for the NL-Pt/YIG and NL-Pt/YIG$^+$ samples, respectively), $d_N$=5nm, and that $G_i \ll G_r$ [63], Eq. (2) yields $G_r$ ~3.2×10$^{13}$ $\Omega^{-1}$m$^{-2}$ and ~4.0×10$^{13}$ $\Omega^{-1}$m$^{-2}$ for the Pt/YIG and Pt/YIG$^+$ interfaces, respectively, which is in very good agreement with our previous results.

The non-local resistance $R_{NL}$ [Fig. 3(c)] shows a $\sin^2(\beta)$ dependence, which is expected for the excitation, transport and detection of magnon spin information through the YIG film [25,29,30]. The physical description of this phenomenon is the following. The current applied in the central Pt wire (injector) produces a transverse spin current (*via* the SHE) that flows along the *z* axis [being **s** parallel to the *y* axis; see Fig. 3(a) for the definition of the axes]. When these spins reach the Pt/YIG interface, they can excite (annihilate) magnons in the YIG film when **s** is parallel (antiparallel) to **M** [25], which produce a change in the magnon population below the Pt injector. These non-equilibrium magnons diffuse through the YIG film and, when they reach the nearby Pt wire (detector), the reciprocal process takes place. Therefore, the non-equilibrium magnons below the Pt detector transform into a non-equilibrium spin imbalance at the Pt/YIG interface, which produces the flow of a pure spin current perpendicular to the interface that is ultimately converted into a perpendicular charge current (along the Pt wire) *via* the ISHE. The combination of all these processes generates the non-local resistance $R_{NL}$ shown in Fig. 3(c) [80].

The angular dependence observed in Fig. 3(c) confirms that the excitation and absorption of propagating magnons in the YIG film are maxima when **s** and **M** are collinear, which occurs for $\beta$=90º and 270º (note that the sign of the signal captured agrees with the sign convention chosen for our experiments [25,29]). Moreover, $V_{NL}$ should be linear with *I* for moderate applied currents [25]. This is confirmed in Fig. 3(c), where it is shown that the same $R_{NL}(\beta)$ curve is obtained for *I*=100 (black) and 300 µA (red). The amplitude of the $R_{NL}(\beta)$ curve measured in our sample is consistent with results reported using YIG films with similar thicknesses [29].

Figure 4 shows the temperature dependence of the amplitude of (a) the SMR and (b) the MST measured in both the NL-Pt/YIG (black squares) and NL-Pt/YIG$^+$ (red circles) samples. The sign of the amplitude of the SMR (local) and the MST (non-local measurements) is indicated with the arrows drawn in Figs. 3(b) and 3(c), respectively. The SMR data is presented normalized to the base resistance, following the same procedure used in the previous case. In Fig. 4(a), we see that the temperature dependence of the SMR in these samples is qualitatively similar to the one observed in the previous experiments [see Figs. 2(c) and 2(d)], which confirms once again the emergence of a surface magnetic frustration in the treated YIG$^+$ substrate at low temperatures.

Interestingly, while the amplitude of the SMR in the temperature range ~150-300 K is only slightly larger in the NL-Pt/YIG$^+$ sample than in the NL-Pt/YIG one (i.e., slight enhancement of $G_r$), the amplitude of the MST is about four times larger [see Fig. 4(b)]. This indicates that in this temperature range the efficiency of the spin-to-magnon conversion (and its reciprocal process) in the treated Pt/YIG$^+$ interface is much higher than in the non-treated Pt/YIG interface, but not related to the change in $G_r$. Instead, it must be associated to the different magnetic properties of the treated YIG$^+$ surface compared to the YIG bulk for temperatures above the emergence of the magnetic frustration. Further studies will be needed in order to fully understand the role of this surface enhancement.

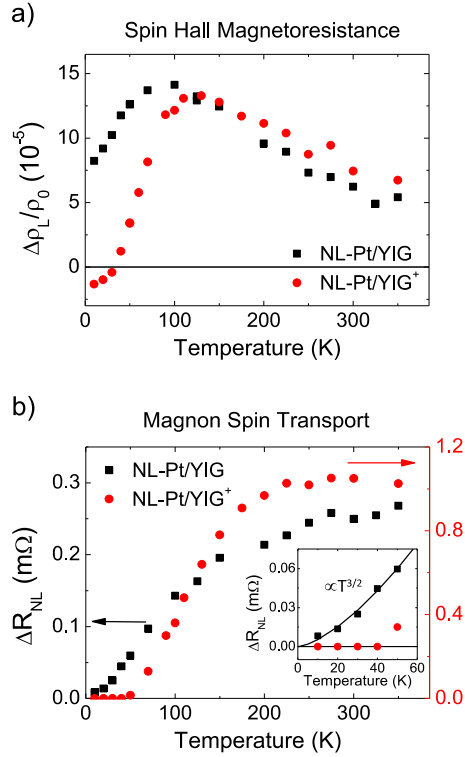

FIG. 4 (color online). Temperature dependence of the amplitude of (a) the SMR and (b) the MST measured in NL-Pt/YIG (black squares) and NL-Pt/YIG$^+$ (red circles). The amplitude is extracted from ADMR measurements performed in the $\beta$ plane at $H$=1 T. Measurements in (a) and (b) are independent of $I$ (at least) to up to 300 μA. The inset in panel (b) shows a zoom of the measured $\Delta R_{NL}$ at low temperatures. Black solid line is a fit to the experimental points to the power law dependence $T^{3/2}$.

The temperature dependence of the amplitude of the MST follows a remarkably different trend than the SMR, which is in agreement with recent reports [29]. In fact, we found that the MST amplitude in the NL-Pt/YIG sample at low temperatures follows a $\sim T^{3/2}$ dependence [see inset in Fig. 4(b)], expected for thermally induced diffusive magnons in the limit of large magnon diffusion lengths (i.e., weak magnon-phonon interactions) [27,29,81,82]. Importantly, the temperature dependence decays more abruptly for the NL-Pt/YIG$^+$ sample, and no MST signal is detected at low temperatures (within the noise level), evidencing that the emergence of the surface magnetic frustration results in the suppression of non-equilibrium diffusive magnons at the surface of the YIG$^+$ film. In other words, the frustrated magnetic surface, which may host a magnon dispersion relation different from the YIG bulk, is preventing the efficient spin-to-magnon conversion (and viceversa) at the Pt/YIG$^+$ interface.

## IV. Conclusions

We demonstrate via SMR and MST measurements in Pt/YIG-based devices that an Ar$^+$-ion milling treatment of the YIG surface has a profound impact in the resulting spintronic phenomena. Beyond a slight increase in the spin-mixing conductance observed for the treated samples at room temperature, which accounts for a better interface quality, we show that the MST is fourfold increased. This elucidates the higher sensitivity of the magnon excitations to fine details in the magnetic properties of the magnetic surface. Moreover, we show that the treated surface of YIG develops a magnetic frustration at low temperature, which makes the SMR signal to reverse the sign below ~45K and dramatically suppresses the spin-to-magnon excitations in these interfaces. Our results give new insights on the interactions between the spins in a NM material with the magnetic moments of a FM at interfaces free from MPE, and show the potential of SMR and MST to explore the magnetic properties of materials with complex magnetic textures and surfaces.

ACKNOWLEDGEMENTS

We thank E. Sagasta for fruitful discussions. This work was supported by the European Union 7th Framework Programme under the European Research Council (257654-SPINTROS) and the Marie Curie Actions (607904-13-SPINOGRAPH), by the Spanish MINECO under Projects No. MAT2012-37638 and MAT2015-65159-R, and by the Basque Government under Project No. PC2015-1-01.